\documentclass[aps,prb,reprint,showpacs,groupedaddress,
footinbib,citeautoscript]{revtex4-1}

\usepackage{graphicx}

\usepackage[utf8]{inputenc}
\usepackage{csquotes}
\usepackage[american]{babel}
\usepackage[T1]{fontenc}
\usepackage{enumerate}
\usepackage{mdwlist}
\usepackage[activate=normal]{pdfcprot}
\usepackage{bbding}
\usepackage{color}
\usepackage{amssymb}
\usepackage{amsmath}
\usepackage{amsfonts}
\usepackage{mathrsfs}

\usepackage{bm}
\usepackage{dcolumn}
\usepackage{color}
\usepackage[colorlinks=true,citecolor=blue]{hyperref}
\hypersetup{colorlinks=true,citecolor=blue,linkcolor=red,urlcolor=blue}

\begin{document}

\title{Spin-dependent Seebeck effect and spin caloritronics in magnetic graphene}
\date{\today}

\author{Babak Zare Rameshti}
\author{Ali G. Moghaddam}
\affiliation{Department of Physics, Institute for Advanced Studies in Basic Sciences (IASBS), Zanjan 45137-66731, Iran}

\begin{abstract}
We investigate the spin-dependent thermoelectric effects in magnetic graphene in both diffusive and ballistic regimes. Employing the Boltzmann and Landauer formalisms we calculate the spin and charge Seebeck coefficients (thermopower) in magnetic graphene varying the spin splitting, temperature, and doping of the junction.
It is found that while in normal graphene the temperature gradient drive a charge current, in the case of magnetic graphene a significant spin current is also established. In particular we show that in the undoped magnetic graphene in which different spin carriers belong to conduction and valence bands, a pure spin current is driven by the temperature gradient. In addition it is revealed that profound thermoelectric effects can be achieved at intermediate easily accessible temperatures when the thermal energy is comparable with Fermi energy $k_BT\lesssim \mu$. By further investigation of the spin-dependent Seebeck effect and a significantly large figure of merit for spin thermopower $\mathcal{Z}_{\rm sp}T$, we suggest magnetic graphene as a promising material for spin-caloritronics studies and applications.  
\end{abstract}

\maketitle
\section{introduction}\label{sect:intro}

Thermoelectric effects, although, known for almost two centuries, have received great attention in recent years due to their crucial relevance in meso- and nanoscopic systems \cite{giazotto2006,dubi2011}. 
Not only can the studies be helpful technologically in managing the generated heat in nanoelectronic devices, however; investigations about thermoelectric effects in mesoscopic regimes are of fundamental interest for condensed matter physicists \cite{giazotto2006}.
Starting in the late 1980's the field of spintronics emerged which focuses on the spin-dependent transport and its coupling to that of charge \cite{Zutic2004,Wolf2001,Awschalom2007}. Along with the fast growing interest in this field, the pioneering work of Johnson and Silsbee showed that, in spintronic and magnetic systems, heat currents can couple to spin currents as well as charge currents \cite{johnson1987}. In recent years some successive unexpected experimental observations of spin Seebeck effects \cite{Uchida2008,Uchida10,Jaworski10,Kirihara12} have attracted a great deal of attention in investigating the thermoelectric and spintronic effects in combination with each other, which has lead to the introduction of a new research field, \emph{spin caloritronics} \cite{Bauer2012,Sinova2010}. Besides many promising applications, some fundamental questions have arisen in this field, particularly about the origin of the spin Seebeck effect in different types of materials varying from metals to insulators. 
\par
Graphene, as a leading material among recently synthesized two-dimensional atomic monolayers, has received a tremendous amount of interest mostly due to its peculiar electronic structure described by the massless Dirac model \cite{neto2009,geim2007}. A large number of possible applications in electronics, optics, nanoscale resonators, and even chemistry were suggested and implemented immediately after its discovery a decade ago. One of the main lines of investigation in graphene from the very beginning has been the electronic transport in a variety of regimes from ballistic to diffusive, and also in the extreme regimes of low density or high magnetic fields \cite{novoselov2005,kim2005}. The experimental observation of 
linear dependence of the conductivity on the carrier density initiated a debate in the theoretical community which guided them to include long-ranged charged scatterers for an adequate description of electron transport (Ref. \onlinecite{das-sarma-2011} and references therein provide a thorough review on this topic). Intriguingly the thermoelectric properties of graphene have been also investigated both theoretically and experimentally with special focus on the neutrality or Dirac point \cite{zuev2009,wei2009,hwang2009,xu2014,sanchez14}.
One of the key findings has been the 
sign change of the thermoelectric power across the charge
neutrality point when the carrier type switches from electron to hole, accompanied by the divergent behavior of the Seebeck coefficient \cite{hwang2009}. 
\par
Besides many other promising applications recently graphene has been suggested for in spintronics devices in particular due to the long spin relaxation
lengths up to a few microns \cite{tombros2007,tombros2008}. Pioneering works of Tombros \emph{et al.} have verified the effective spin injection into graphene via nonlocal magnetoresistance measurements. In addition it has been suggested that spin qubits based on graphene can be used as building blocks for quantum computing \cite{trauzettel2007}.
Interestingly a variety of methods have been
suggested to create magnetic graphene, 
besides some theoretical predictions about intrinsic ferromagnetism in it \cite{peres2005,Son2006}. In practice one can use 
an insulating ferromagnetic substrate or, alternatively, add a magnetic material or magnetic impurities on top of the graphene sheet to induce spin imbalance (for a review on magnetism in graphene, see Ref. \onlinecite{yazyev}). In addition very recently the proximity-induced ferromagnetism in graphene/YIG heterostructure has been revealed which indicates a large exchange interaction \cite{shi15}. In contrast to common magnetic materials due
to the gapless excitation spectrum of graphene, and the fine
tunability of its chemical potential $\mu$, the spin-splitting energy between the up- and down-spin carriers can be even comparable with $\mu$. So there exists a regime in which majority
and minority spins belong to different bands, conduction and valence. We have called this phase as spin-chiral due to the coupling of the real spin and the chirality and already some of its transport characteristics have been explored \cite{moghaddam2010,zareyan2008,moghaddam2008,zare2013prb,
zare2013}. We should note that based on Zeeman splitting such
a spin-chiral graphene has been experimentally realized which shows the spin Hall effect without spin-orbit interaction \cite{Abanin2011}. 
\par
\begin{figure}[tp]
 \includegraphics[width=0.35 \linewidth]{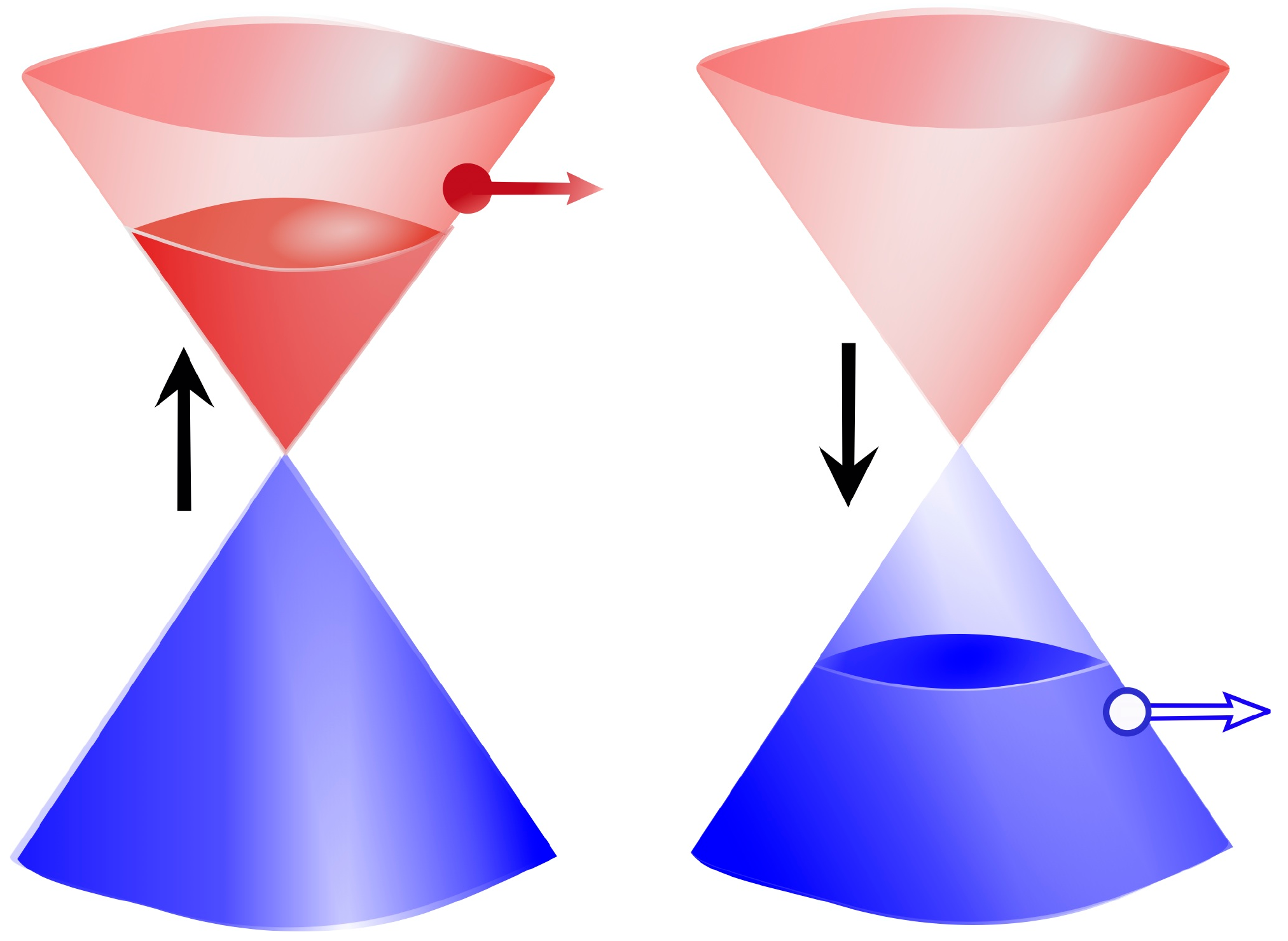}
\hspace{1.5 cm} \includegraphics[width=0.35 \linewidth]{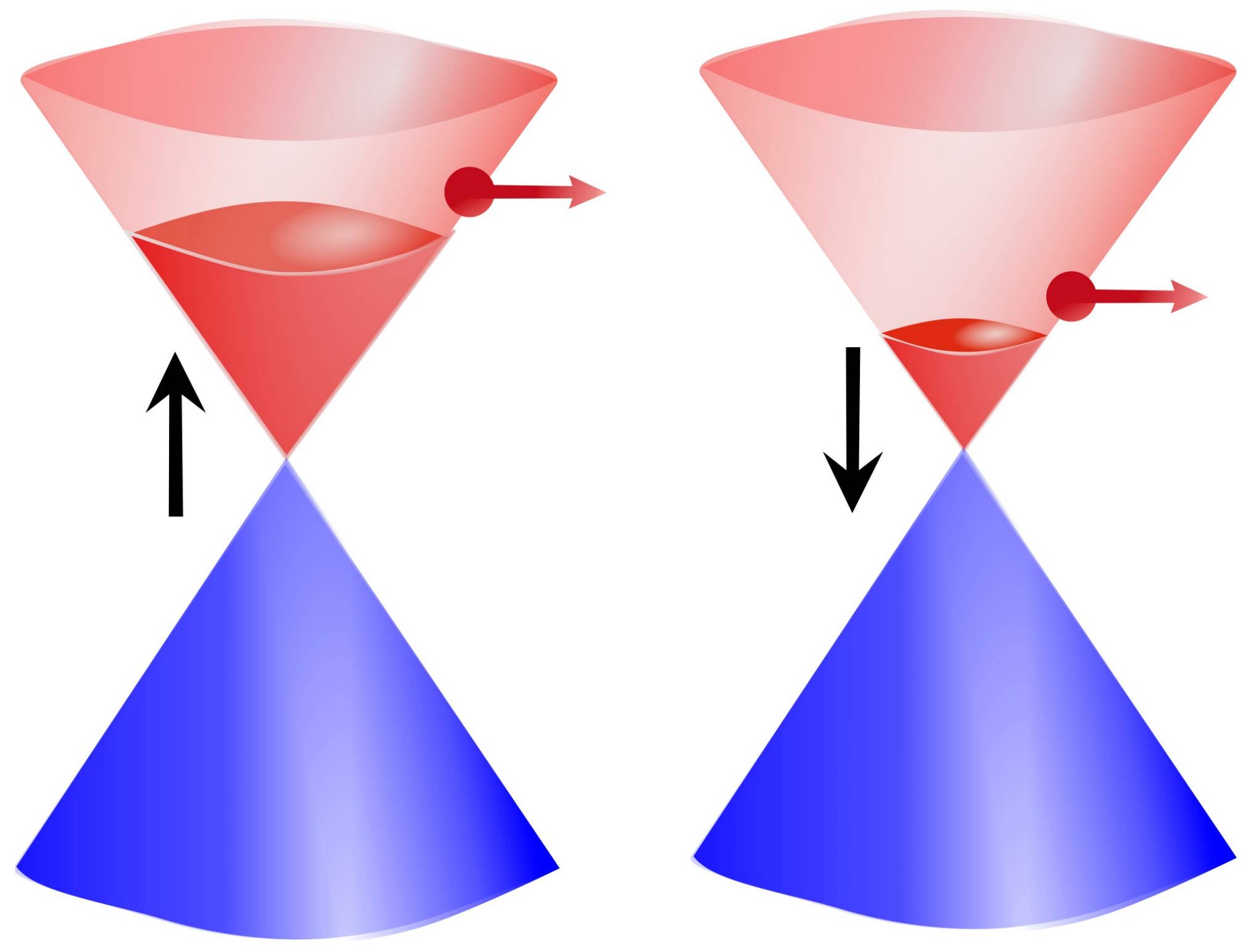}
  \caption{(Color online) The spin band structure of magnetic graphene for two regimes of $V_{\rm ex}>\mu$ (left panel) and  $V_{\rm ex}<\mu$ (right panel). Small filled (empty) circles and the arrow attached to them show the thermally excited electrons (holes) and their velocity directions.   }
\label{fig1}
\end{figure}

In this paper we investigate the combination of charge, heat, and spin transport in graphene in the context of spin caloritronics and spin-dependent thermoelectric phenomena. We consider a magnetic graphene sheet in both ballistic and diffusive regimes when thermal gradients and bias voltages are applied to it. Employing the Landauer-B\"uttiker scattering method and Boltzmann transport equation for the two regimes we obtain the spin-dependent Seebeck and Peltier coefficients and the spin-dependent figure of merit which is a measure of thermoelectric efficiency. Our findings show that while in the absence of exchange splitting the temperature gradients drive only a charge current, in the case of magnetic graphene a spin current is also established which can be very large in comparison with charge current. In fact our key finding is for the case of undoped spin-chiral magnetic graphene in which different spin carriers are electrons and holes having the same density (see left panel of Fig. \ref{fig1}): a pure spin current (without charge current) is driven by the imposed temperature gradient.
An explanation of this effect can be provided noting that the temperature gradient in spin-chiral graphene drives electrons from up-spin subband and holes from down-spin subband. Remembering the fact that holes carry opposite spin and charge of the corresponding electron, both types of carriers (electron and holes) carry the same spin but opposite charges which leads to the pure spin current. However for weakly magnetized and doped graphene when both spins belongs to the conduction or valence bands (right panel in Fig. \ref{fig1}), both spin and charge currents exist while the second dominates the first one.
Adding the facts that spin relaxation is very weak in graphene and its electronic properties can be easily tuned,
these results show that magnetic graphene could be promising for the spin-caloritronic applications rather than common magnetic metals.

\section{Theoretical model and basic formalism}
In order to study the spin and charge thermoelectric properties of magnetic graphene we work in the linear response regime where the relation between driving forces and the resulting generalized currents are linear. In a thermoelectric device the driving forces are temperature gradient $\nabla_r T$, external electric field $E$, and density gradients $\nabla_r n$ where the two last ones can be combined in an effective field $\mathcal{E}=E+\frac{1}{e}\frac{\partial\mu}{\partial n}\nabla_{r}n$ with $\mu$ indicating the chemical potential.
These fields can drive carriers leading to the charge and heat currents which can differ for different spin channels in general. It is believed that in graphene spin relaxation time is so long that in mesoscopic samples we can treat the two spin channels almost independently. As a result the charge and heat currents carried by spin $s$ electrons which can be denoted by $j_{s} $ and $j^q_{s}$, respectively, are linearly related to the effective electric field $\mathcal{E}=E+\frac{1}{e}\frac{\partial\mu}{\partial n}\nabla_{r}n$ and the temperature gradient as
\begin{eqnarray}
\begin{pmatrix}
j_{s} \\ j^q_{s}
\end{pmatrix} =\begin{pmatrix}
L_{s}^{11} & L_{s}^{12} \\L_{s}^{21} & L_{s}^{22} \end{pmatrix}\begin{pmatrix}
\mathcal{E} \\ -\nabla T
\end{pmatrix}
\label{LL}
\end{eqnarray}
By definition the first component $L^{11}_s$
is the conductance $G$ and the two off-diagonal components are thermoelectric coefficients which are related to each other with Onsager relation ($L^{21}_s=T L^{12}_s$). The last component $L^{22}_s$ contributes in the $s$-electron 
thermal conductivity defined by
\begin{equation}
\mathcal{K}_{s}=\frac{L_s^{12}L_s^{21}-L_s^{11}L_s^{22}}{L_s^{11}}
\label{kappa_s}
\end{equation}
In the upcoming subsections we will give the explicit relations for the matrix elements $L^{ij}_s$ in the diffusive and ballistic regimes. 
\par
The spin-dependent Seebeck and the Peltier coefficients 
which, for each spin channel, describe the voltage generation due to the temperature gradient and heat current induction due to the charge current, respectively, are then given by
\begin{equation}
\mathcal{S}_{s}=\frac{L_{s}^{12}}{L_{s}^{11}}~~,~~~~
\Pi_{s}=\frac{L_{s}^{21}}{L_{s}^{11}}.\label{sandpi}
\end{equation}
From these relations one can define the charge and spin Seebeck and the Peltier coefficients which are as follows,
\begin{eqnarray}
&&\mathcal{S}_{\rm ch}=(\mathcal{S}_{\uparrow}+\mathcal{S}_{\downarrow})/2~~~~,~~
\mathcal{S}_{\rm sp}=\mathcal{S}_{\uparrow}-\mathcal{S}_{\downarrow}\nonumber\\
&&\Pi_{\rm ch}=(\Pi_{\uparrow}+\Pi_{\downarrow})/2~~~,~~
\Pi_{\rm sp}=\Pi_{\uparrow}-\Pi_{\downarrow}
\end{eqnarray}
\par
The ability of a material to efficiently produce thermoelectric power is usually described by a dimensionless figure of merit denoted by $\mathcal{Z}T$. In spin caloritronics we can generalize this concept for the resulting charge and spin currents due to the temperature gradients separately. So the charge and spin figures of merit for a magnetic system can be defined versus Seebeck coefficients as
\begin{eqnarray}
\mathcal{Z}_{\rm ch(sp)}T=\frac{\sigma_{\rm ch(sp)}\mathcal{S}_{\rm ch(sp)}^{2}T}{\mathcal{K}}.
\end{eqnarray}
Here, $\sigma_{\rm ch}=\sigma_{\uparrow}+\sigma_{\downarrow}$ ($\sigma_{\rm sp}=\vert\sigma_{\uparrow}-\sigma_{\downarrow}\vert$) denotes the charge (spin) conductivity of the system and the   
the electron thermal conductivity is given by $\mathcal{K}=\mathcal{K}_{\uparrow}+\mathcal{K}_{\downarrow}$.
We concentrate on low enough temperatures where only electrons contribute effectively in thermal transport. 
In Sec. \ref{experi} based on some estimations, we will discuss how this assumption is verified and what are its limitations.
\par
In the remaining of this section, the theoretical frameworks to calculate the spin and charge thermoelectric coefficients in the diffusive and ballistic transport regimes employing Boltzmann and Landauer formalisms, respectively, will be presented. 

\subsection{Diffusive regime: Boltzmann transport}

In this section we give the semiclassical Boltzmann equation to establish the transport coefficients in the diffusive regime. In particular, we take into account two important cases of short-range (SR) impurities with Dirac delta potentials and long-range (LR) Coulomb impurities in our investigation. The spin-dependent thermoelectric properties due to the presence of both electric fields and temperature gradient will be found in the scheme of relaxation time approximation.
\par
In the diffusive regime, the transport coefficients can be calculated from the following general expression for electron current and energy flux density,
\begin{equation}
\left[ \begin{array}{c} {\bf j}_s \\{\bf j}_s^q \end{array}\right]=\int\frac{d^{2}k}{(2\pi)^{2}}\left[\begin{array}{c}
-e \\ \varepsilon_s(k)-\mu
\end{array}\right] {\bf v}_s(k)g_s(k)\label{j-jq}
\end{equation}
in which ${\bf v}_s(k)$ is the semiclassical velocity of the spin $s$ carriers. The nonequilibrium distribution function $g_s(k)$ describes the evolution of the electron distribution in the presence of external perturbations. The Boltzmann formalism in the relaxation time approximation scheme and in the linear response leads to the following expression for disturbed function $g_s(k)$ as
\begin{eqnarray}
g_s(k)=\tau_s(k)\left(\frac{\partial f_{0}}{\partial\varepsilon}\right) {\bf v}_s(k)\cdot\left[e{\mathcal E}+  \frac{\varepsilon_s(k)-\mu}{T}\nabla T\right] \label{noneq-g}
\end{eqnarray}
with $\tau_s(k)$ denoting the spin-dependent relaxation time and $f_0(\varepsilon)$ the equilibrium-state Fermi-Dirac distribution at temperature $T$. We note that since graphene has an isotropic dispersion relation, the relaxation time $\tau$ depends only on the energy of electrons $\varepsilon$. 
\par
By invoking the above expression for $g_s(k)$ into Eq. (\ref{j-jq}) for spin-dependent charge and heat currents, 
the matrix coefficients $L_{s}^{ij}$ can be expressed in terms of some spin-dependent \emph{kinetic coefficients} $\mathcal{L}_{s}^{\alpha}$ as the following,
\begin{eqnarray}
\begin{pmatrix}
L_{s}^{11} & L_{s}^{12} \\L_{s}^{21} & L_{s}^{22} \end{pmatrix}=
\begin{pmatrix}
\mathcal{L}_{s}^{0} & -\mathcal{L}_{s}^{1}/eT \\ \mathcal{L}_{s}^{1}/e & -\mathcal{L}_{s}^{2}/e^{2}T 
\end{pmatrix}.
\label{LL-ells}
\end{eqnarray}
All of the coefficients obey the relation
\begin{equation}
\mathcal{L}_{s}^{\alpha}=\int d\varepsilon\left(-\partial f_{0}/\partial\varepsilon\right)(\varepsilon-\mu)^{\alpha}\sigma_{s}(\varepsilon),\label{ells}
\end{equation}
with spin-dependent conductivity given by
\begin{eqnarray}
\sigma_{s}(\varepsilon)&=&e^{2}\tau_{s}(\varepsilon)\int\frac{d^{2}k}{(2\pi)^{2}}\delta[\varepsilon-\varepsilon_s(k)]v_s(k)v_s(k)\nonumber\\
&=&e^{2}v^2_s(\varepsilon)\tau_{s}(\varepsilon)\rho_s(\varepsilon),
\label{sigma_s}
\end{eqnarray}
with spin-dependent density of states (SDOS) $\rho_s(\varepsilon)$.
The formalism introduced so far is general for any isotropic magnetic material and all of the thermoelectric properties described by $L_s^{ij}$ can be found as functions of spin-dependent relaxation time $\tau_s$ and SDOS $\rho_s$. Now we switch to the case of our investigation, magnetic graphene.
\par
A monolayer graphene sheet at the presence of induced spin splitting can be described by a low-energy Dirac Hamiltonian of the form
\begin{equation}
\mathcal{H}=\hbar v_{F}\hat{s}_{0}\otimes\hat{\bm{\sigma}}\cdot{\bf p}-V_{\rm ex}\hat{s}_{z}\otimes\hat{\sigma}_{0}\label{dirac}
\end{equation}
with Fermi velocity $v_{F}$, momentum ${\bf p}=(p_{x},p_{y})$, and exchange splitting $V_{\rm ex}$. Pauli matrices $\hat{\sigma}_i$ and $\hat{s}_i$ ($i=0,\cdots,3$) operate on the subspaces of pseudospin (originating from two different trigonal sublattices $A$ and $B$ of the graphene's hexagonal structure) and real spin, respectively. The spin-dependent band dispersion then follows $\varepsilon_{s}(k)=\alpha \hbar v_{F}k -sV_{\rm ex}$ with $s=\pm 1$ corresponding to the two spin directions and $\alpha=\pm 1$ indicating the chirality of states. Since the velocity of carriers in graphene is constant and the SDOS is given by $\rho_{s}(\varepsilon)=\vert\varepsilon+sV_{\rm ex}\vert/\pi(\hbar v_{F})^{2}$ the spin-dependent Boltzmann conductivity of magnetic graphene takes the Drude form as
\begin{equation}
\sigma_s(\varepsilon)=\frac{e^2}{h}\frac{|\varepsilon+sV_{\rm ex}|\tau_s(\varepsilon)}{\hbar}.
\label{Dirac-conductivity}
\end{equation}
Early investigations of quantum transport in graphene at the presence of impurities have shown that the relaxation time for the SR impurities varies inversely with the DOS as $\tau(\varepsilon) \propto 1/\rho(\varepsilon)$ while the LR Coulomb impurities result in $\tau(\varepsilon) \propto \rho(\varepsilon)$ \cite{das-sarma-2011,nomura2007}. Therefore in magnetic graphene, the conductivity becomes constant when only SR scatterers are present while the conductivity caused by scattering from LR impurities is proportional to the square of density of states as $\sigma_{s}(\varepsilon)\propto(\varepsilon+sV_{\rm ex})^{2}$. In the next section we will use the relations of conductivities to obtain the Seebeck coefficients and corresponding figures of merit. 

\subsection{Ballistic regime: Landauer-B\"uttiker formula}

Within the Landauer-B\"{u}ttiker approach the electric and thermal currents carried by electrons with spin $s$ are obtained from the transmission probabilities $T_{s}(\varepsilon,\phi)$ integrated over the energy $\varepsilon$ and the angle $\phi$,
\begin{eqnarray}
\left[ \begin{array}{c} {I}_s \\{I}_s^q \end{array}\right]&=&\frac{W}{\pi^2 \hbar}\int_{-\pi/2}^{\pi/2}d\phi \, \cos\phi \int_{-\infty}^{\infty}d\varepsilon ~\left[\begin{array}{c}
-e \\ \varepsilon-\mu
\end{array}\right] \nonumber\\
&&\qquad\times~ \rho_{s}(\varepsilon) T_{s}(\varepsilon,\phi)\left[f_{L}(\varepsilon)-f_{R}(\varepsilon)\right]\label{j-jq-bal}
\end{eqnarray}
where $W$ is the sample width and $f_{L}(\varepsilon)$ and $f_{R}(\varepsilon)$ are the Fermi-Dirac distribution functions of the left and right electronic leads, respectively. 
Assuming the linear response regime, we can expand the difference of Fermi-Dirac functions in the above formula up to linear terms in a small bias voltage $V$ and temperature difference $\Delta T$ between two reservoirs. This results in a relation very similar to Eq. (\ref{LL}) with thermoelectric conductances $L_s^{ij}$ related to the kinetic coefficients according to Eq. (\ref{LL-ells}). These coefficients for the ballistic transport regime are obtained after some straightforward algebra,
\begin{eqnarray}
\mathcal{L}_{s}^{\alpha}&=&
G_{0}\int_{-\pi/2}^{\pi/2}d\phi \cos\phi\nonumber\\ 
&&\times \int_{-\infty}^{\infty}d\varepsilon 
(-\frac{\partial f}{\partial\varepsilon})
(\varepsilon-\mu)^{\alpha}\rho_{s}(\varepsilon)T_{s}(\varepsilon,\phi),\label{ells-bal}
\end{eqnarray}
in which $G_0=(e^2/\hbar)(W/\pi^2)$ denotes the ideal conductance of the junction. It is easy to note that this relation is in complete accordance with Eq. (\ref{ells}) in which only the conductivity is replaced with the conductance of the ballistic system given by the Landauer-B\"uttiker formula. So the only thing we need is to determine the transmission probabilities $T_{s}(\varepsilon,\phi)$.

In the ballistic regime we consider a magnetic graphene sheet between two electrodes in which their chemical potentials despite a very small bias are kept at $\mu_0$ while the central region's $\mu_m$ doping can be varied. Diagonalizing the Hamiltonian (\ref{dirac}) results in eigenstates $\psi_{{\bf k}\alpha s}=\vert s\rangle\otimes ( \alpha e^{i\phi_{\bf k}},1)^T$, where $\alpha=\text{sign}(\mu+sV_{\rm ex})$ indicates the band index and $\phi_{k}=\arctan(k_{y}/k_{x})$ specifies the propagation direction. The total scattering wave function, corresponding to an incoming excitation coming from the left reservoir, inside each region, can be written as
\begin{eqnarray}
\psi_s^{L}&=&e^{ik_{s}x}\psi_{(k_s,k_y)\alpha s}+r_{s}e^{-ik_{s}x}\psi_{(-k_s,k_y)\alpha s}\nonumber\\
\psi_s^{m}&=&a_s^{+}e^{iq_{s}x}\psi_{(q_s,k_y) \alpha' s}+a_s^{-}e^{-iq_{s}x}\psi_{(-q_s,k_y)\alpha' s}\nonumber\\
\psi_s^{R}&=&t_{s}e^{ik_{s}x}\psi_{(k_s,k_y) \alpha s}\label{scatter}
\end{eqnarray}
with $q_{s}=\sqrt{(\varepsilon+\mu_{m}+sV_{\rm ex})^{2}-(\hbar v_{F}k_{y})^{2}}/\hbar v_{F}$ and $k_s=\sqrt{(\varepsilon+\mu_{0}+sV_{\rm ex})^{2}-(\hbar v_{F}k_{y})^{2}}/\hbar v_{F}$ (for the sake of simplicity the overall factor $e^{ik_yy}$ is dropped in all components of wave functions). The transmission $t_{s}$ and $r_{s}$ reflection amplitudes are determined by matching the wavefunctions at the interfaces $x=0$ and $x=L$ and subsequently used to calculate the total transmission probability $\vert t_{s}\vert^{2}$ as
\begin{eqnarray}
&&T_{s}(\varepsilon,\phi_{k})\nonumber\\
&&~~~~=\frac{1}{\cos^{2}(q_{s}L)+\sin^{2}(q_{s}L)\left(\frac{1-\alpha'\sin\phi_{k}\sin\phi_{q}}{\cos\phi_{k}\cos\phi_{q}}\right)^{2}}\quad\label{trans}
\end{eqnarray}
We recall that there is no mechanism that couples states with opposite spin indices so no off-diagonal spin channel mixing terms will appear in our calculations.
\section{Results and Discussion}
In this section our numerical and analytical results will be presented. We focus on the charge and spin Seebeck coefficients ($\mathcal{S}_{\rm ch}$, $\mathcal{S}_{\rm sp}$) and their corresponding figures of merit $\mathcal{Z}_{\rm ch}T$ and $\mathcal{Z}_{\rm sp}T$. We can also find the Peltier coefficients
$\Pi_{\rm ch}$ and $\Pi_{\rm sp}$ in both diffusive and ballistic regimes. However according to the so-called Thomson relation $\Pi=T\mathcal{S}$, which is originated from the symmetry properties of the coefficients $L^{ij}_s$ demanded by Onsager reciprocity, there is no new information on the Peltier coefficients. We divide this section into two parts concentrating on the diffusive and ballistic systems, respectively.
\subsection{Diffusive transport}
It is well known that electron-hole asymmetry around the Fermi level in the band structure or transport properties is responsible for the thermoelectric effects. In fact the key role in thermoelectric effects is played by $\mathcal{L}_s^{(1)}$ which according to Eq. (\ref{ells}) vanishes when $\sigma(\varepsilon)$ is a symmetric (even) function of $\varepsilon-\mu$. In the case of graphene, at very first glance, the Dirac dispersion relation and linear energy dependence of DOS suggest a possible source of asymmetry in $\sigma(\varepsilon)$ away from the neutrality point which can lead to thermoelectric phenomena. However as we have seen in the previous section when only short range scatterers are present the conductivities $\sigma_s(\varepsilon)$ become constant. Therefore diffusive transport caused by SR impurities leaves magnetic graphene with no thermoelectric effects with vanishing charge and spin Seebeck and Peltier coefficients. 
\par
In contrast at the presence of long range Coulomb impurities, which are in fact the dominant scatterers in most graphene samples, the spin-dependent conductivities have explicit energy dependence. Invoking the quadratic energy dependence of conductivities $\sigma_s$ in Eq. (\ref{ells}) and performing the integrations over energies we find a simple form for the Seebeck coefficient of spin-$s$ carriers, 
\begin{equation}
{\cal S}_{s}=-\frac{k_B}{e}  \frac{ 2 k_BT (\mu+sV_{\rm ex})}{ (3/\pi^2)(\mu+sV_{\rm ex})^2+(k_BT)^2}.
\label{s-s-analytic}
\end{equation}
in which $\mu=\mu(T,V_{\rm ex})$ depends explicitly on temperature and exchange splitting. The method of the calculation of $\mu(T,V_{\rm ex})$ and the behavior of chemical potential as a function of exchange and temperature will be presented in the Appendix. 
As one can see from Eq. (\ref{s-s-analytic}) two spin-dependent Seebeck coefficients ${\cal S}_{s}$ reach their maximum absolute values $(\pi/\sqrt{3})(k_B/e)$ at temperatures $k_BT=(\sqrt{3}/\pi)(\mu+s V_{\rm ex})$, respectively. In addition we see that each of the coefficients ${\cal S}_{s}$ passes from zero and changes sign when the Fermi level of the corresponding spin subband lies at the Dirac point $\mu+sV_{\rm ex}=0$. This is similar to the well-known effect in semiconductors in which the thermopower for $n$ and $p$ types has opposite sign and based on this effect devices made of $p-n$ junctions are used for electronic cooling. However a big advantage in the case of graphene is provided by the fine-tunability of doping in it. So in real experimental situations one can play with $\mu_0$, exchange splitting, and also temperature to cover a wide range of parameter space.
\par
\begin{figure}[tp]
\includegraphics[width=0.95\linewidth]{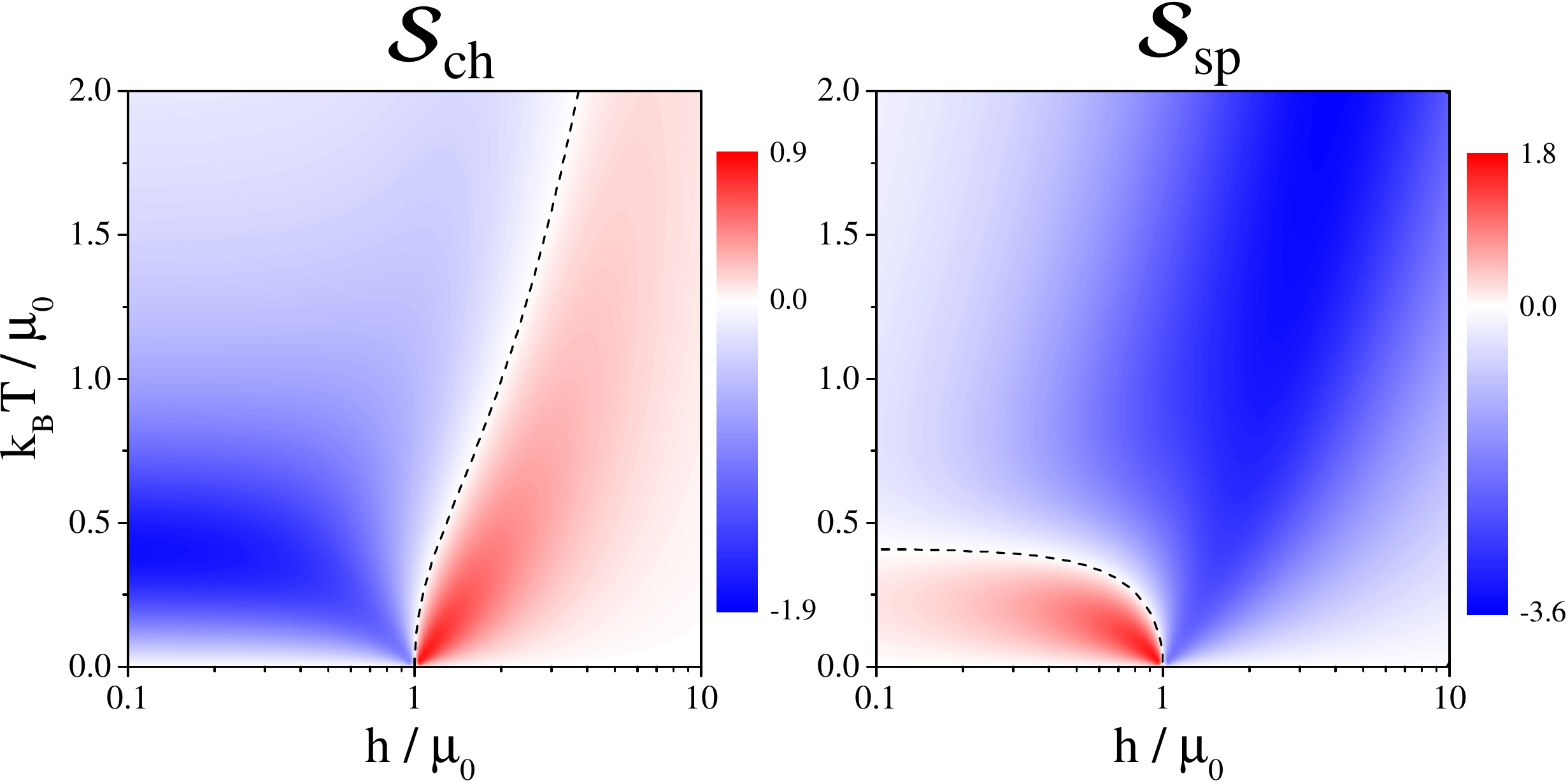}
\caption{\label{fig2}
(Color online) (a) Charge and (b) spin Seebeck coefficients of magnetic graphene as functions of normalized exchange field and temperature $k_{B}T/\mu_{0}$ at the presence of long-range Coulomb impurities. The dashed lines indicate the curves in which the Seebeck coefficients vanish. The scale of Seebeck coefficients in all plots is $k_B/e$.}
\end{figure}

Now we turn the discussion to ${\cal S}_{\rm ch,sp}$ which are more feasible quantities in real experiments. 
Using  Eq. (\ref{s-s-analytic}) charge and spin Seebeck coefficients can be easily obtained, 
\begin{eqnarray}
\label{s-ch}
{\cal S}_{\rm ch}&=&-\frac{k_B}{e} \, \frac{2 k_BT \,\mu  [\frac{3}{\pi^2}(\mu^2-V_{\rm ex}^2) +(k_BT)^2] }{\prod_{s} [ \frac{3}{\pi^2} (\mu+sV_{\rm ex})^2+(k_BT)^2 ]   },\\
\label{s-sp}
{\cal S}_{\rm sp}&=&-\frac{k_B}{e} \, \frac{4 k_BT \, V_{\rm ex}  [\frac{3}{\pi^2}(-\mu^2+V_{\rm ex}^2) +(k_BT)^2] }{\prod_{s} [ \frac{3}{\pi^2} (\mu+sV_{\rm ex})^2+(k_BT)^2 ]   }.~~~
\end{eqnarray}
Inserting the numerically calculated $\mu(T,V_{\rm ex})$ in the above relations, the variation of thermopowers ${\cal S}_{\rm ch,sp}$ with temperature and spin splitting is obtained as shown in Fig. \ref{fig2}. As we expect at very low temperatures, $k_{B}T \ll \mu_0$, charge and spin Seebeck effects are very weak and go to zero linearly with $k_{B}T$. On the other hand at some intermediate temperatures when the thermal energy $k_{B}T$ is comparable with the spin-dependent Fermi levels measured from the Dirac points ($\mu\pm V_{\rm ex}$) profound Seebeck effects can be observed. 
\par
When the spin splitting is small ($V_{\rm ex}\lesssim \mu_0$) both up- and down-spin Fermi levels lie in the conduction band and thermally activated electrons of both spins move along temperature gradient which result in a charge accumulation gradient in the opposite direction due to the negative charge of the electrons. Therefore a negative charge thermopower is obtained for ($V_{\rm ex}\lesssim \mu_0$). 
By further increase in the exchange splitting $V_{\rm ex}\gtrsim \mu_0$ then one of the spin subbands' Fermi level goes to the valence band and then the holes from the spin-down subband will be thermally activated. Such excitations carry positive charge current and as a result their contribution in the charge Seebeck effect has positive sign while spin up electrons from the conduction band still have a negative contribution, which means the excitations from two spin subbands compete with each other. By further increase of the exchange the contribution of minority spin carriers from the hole band dominates and as one can clearly see from Eq. (\ref{s-ch}) at $V_{\rm ex}=\sqrt{\mu^2+(\pi k_BT)^2/3}$ (indicated by the dashed line in Fig. \ref{fig2} ) the charge thermopower changes its sign. 
\begin{figure}[tp]
\includegraphics[width=0.95\linewidth]{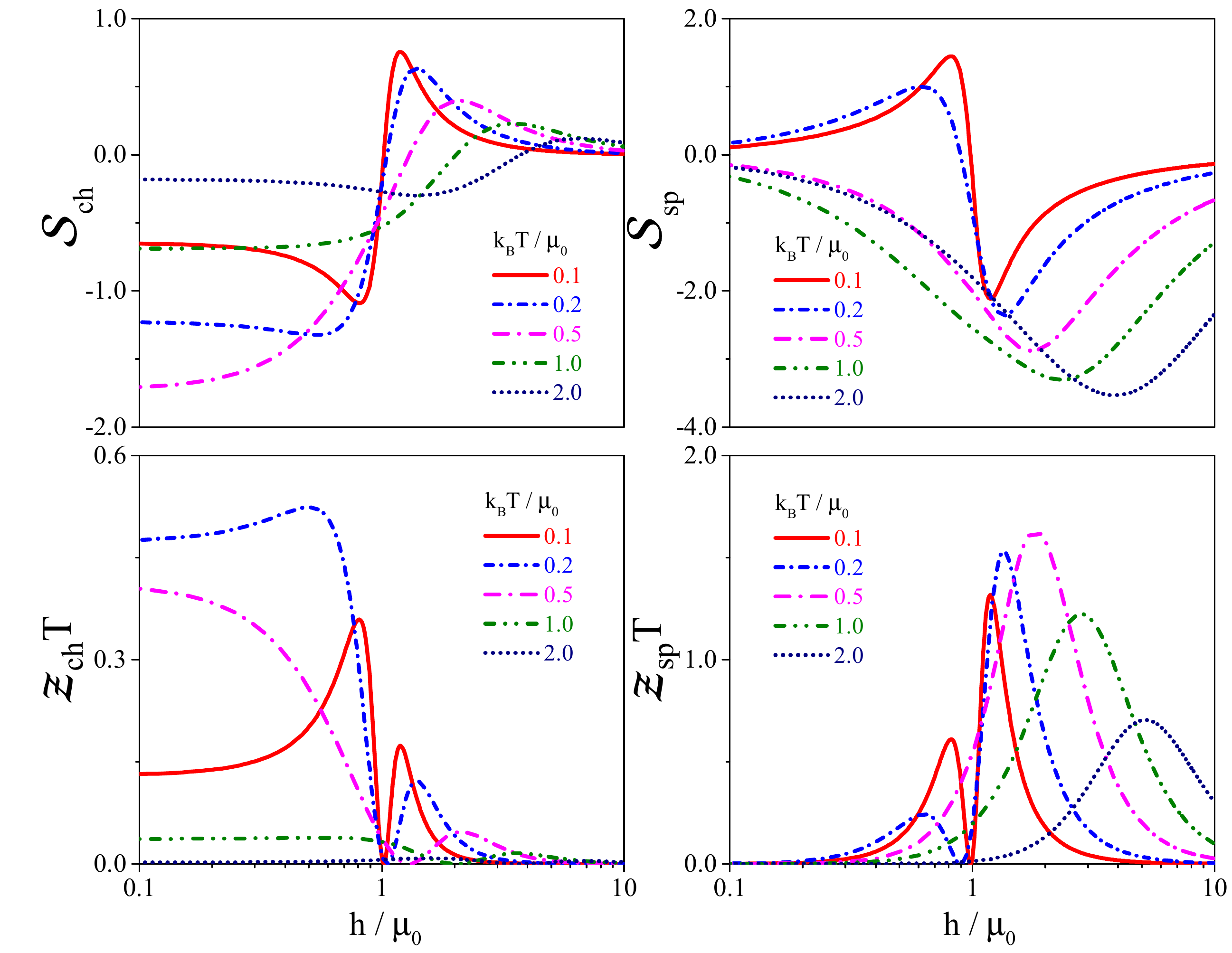}
\caption{(Color online) Charge and spin thermopower and corresponding figures of merit are given as functions of normalized exchange splitting for different dimensionless temperatures $k_{B}T/\mu_{0}$.}\label{fig3}
\end{figure}
\par
The spin Seebeck effect behaves in a somehow opposite way with the variation of exchange field. At low $V_{\rm ex}$ electrons carrying different spins compete with each other to result in a spin accumulation caused by the temperature gradient. As we see in the right panel of Fig. \ref{fig2} for not so high temperatures $k_BT/\mu_0$ the minority spins are dominant and as a result unlike $S_{\rm ch}$ a positive spin Seebeck effect is observed. But at higher temperatures ${\cal S}_{\rm sp}$ becomes all negative dominated by majority up spins. Upon increasing the exchange splitting when the down spins' Fermi level goes to the valence band both up-spin electrons and down-spin holes, which carry the same intrinsic angular momentum, accompany each other to give a strong spin signal. In fact as one can immediately see from Eq. (\ref{s-sp}) for when $V_{\rm ex}^2=\mu^2- (\pi k_BT)^2/3$ the sign of the spin thermopower changes and for higher exchanges $S_{\rm sp}$ becomes negative.  
\par
Now the key finding of our work is the fact that by moving along the curve $V_{\rm ex}^2=\mu^2+(\pi k_BT)^2/3$, we can completely turn off the charge Seebeck effect, while a spin Seebeck effect can be observed. This is clear if we compare two plots in Fig. \ref{fig2} and notice that along the line of $S_{\rm ch}=0$ a large negative spin Seebeck coefficient is obtained which is given by
\begin{eqnarray}
{\cal S}_{\rm sp}=-\frac{k_B}{e} \frac{2\pi}{\sqrt{3}}\sqrt{1-\left(\frac{\mu}{V_{\rm ex}}\right)^{2}}.
\end{eqnarray}
As a result a maximum value $|{\cal S}_{\rm sp}|=(2\pi/\sqrt{3}) k_B/e$ can be reached in the absence of corresponding charge signal.
\begin{figure}[tp]
\includegraphics[width=0.95\linewidth]{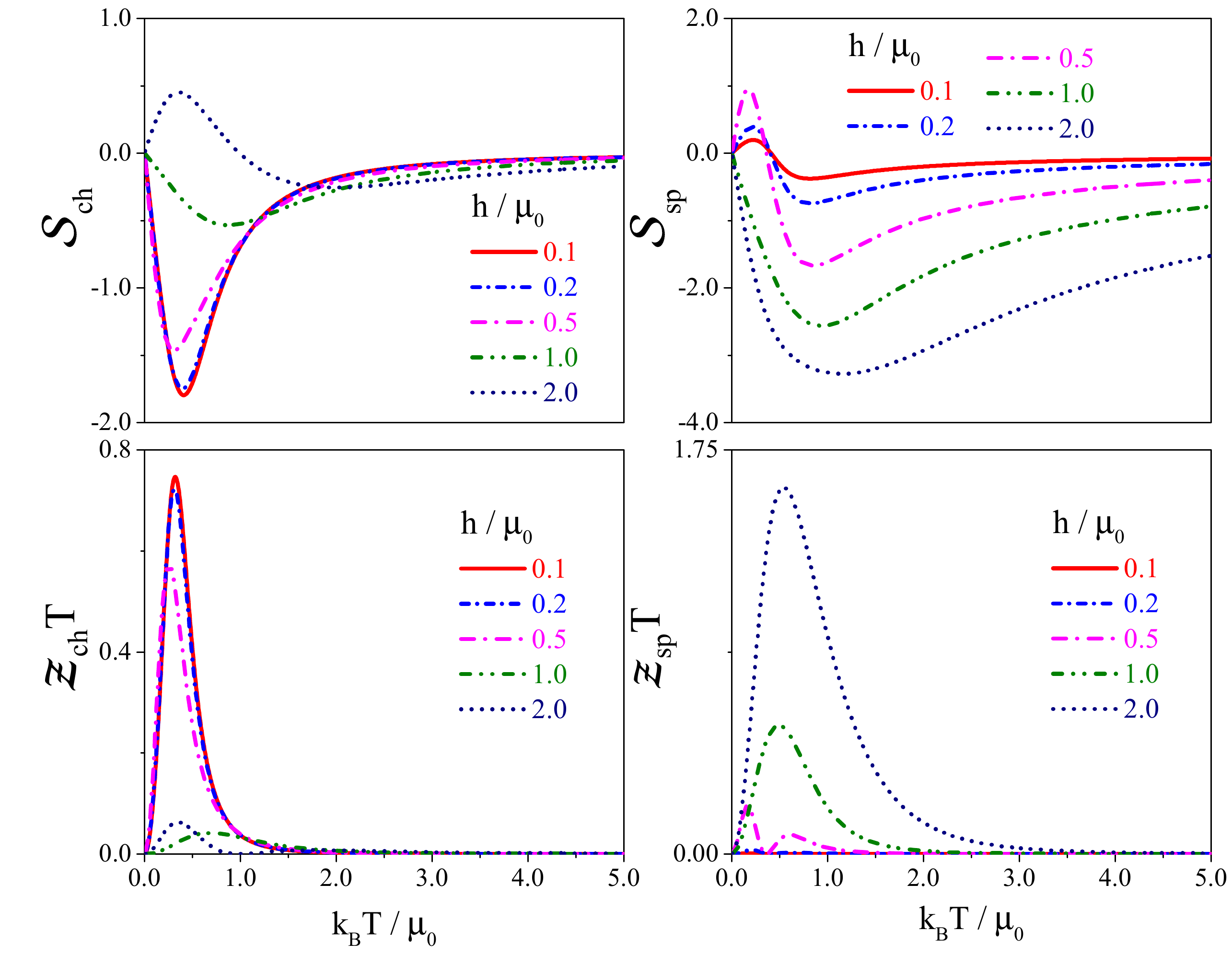}
\caption{\label{fig4}
(Color online) The variations of thermopowers and corresponding figures of merit are given but as functions of dimensionless temperature $k_{B}T/\mu_{0}$ for different exchange splittings $V_{\rm ex}/\mu_0$.}
\end{figure}
\par
In order to see the exchange dependence of the thermopowers more clearly, Fig. \ref{fig3} shows the charge and spin Seebeck coefficients and the corresponding figures of merit as functions of $V_{\rm ex}/\mu_0$ for some different temperatures. We see that although charge thermopower always shows sign change at the vicinity of $V_{\rm ex}\sim \mu_0$, spin thermopower becomes all negative at higher temperatures, irrespective of spin splitting strength $V_{\rm ex}$. This is again a clear manifestation of the possibility of pure spin current caused by temperature gradient.
In fact at higher $V_{\rm ex}$ this is easily understood from the fact that conduction band spin-up electrons and valence band spin-down holes accompany each other to give rise to a negative ${\cal S}_{\rm sp}$. At lower $V_{\rm ex}$ as we mentioned above by increasing temperature majority up spins dominate the thermoelectric effect and since they carry negative current, the spin Seebeck coefficient remains still negative. In addition as expected the spin (charge) figure of merit reaches its maximum value for some splitting above (below) the chemical potential $\mu_0$. The figures of merit for both spin and charge Seebeck effects becomes large (of the order of 1) at some intermediate temperatures $k_{B}T\lesssim \mu_0$ where the thermoelectric effect is very strong while the heat transport is not. On the other hand at higher temperatures the thermopowers decrease as the inverse of $T$ and subsequently the figures of merit show decline with temperature. These effects can be seen from Fig. \ref{fig4} where the variations of ${\cal S}_{\rm ch,sp}$ and ${\cal Z}_{\rm ch,sp}T$ are shown with temperature $k_BT/\mu_0$. These results again clearly show that the strong thermoelectric effects can be seen at the intermediate temperatures when $k_BT$ is comparable with Fermi levels $\mu\pm V_{\rm ex}$ measured from neutrality point. It is worth noting again that in graphene the doping can be varied easily and as a result one can even reach the regimes in which the Fermi energy is comparable to thermal energy $k_BT$. So unlike conventional metals with very large Fermi energy in comparison with $k_BT$, it is reasonable to reach the most efficient values for thermoelectric responses.    
\begin{figure}[tp]
\includegraphics[width=0.95\linewidth]{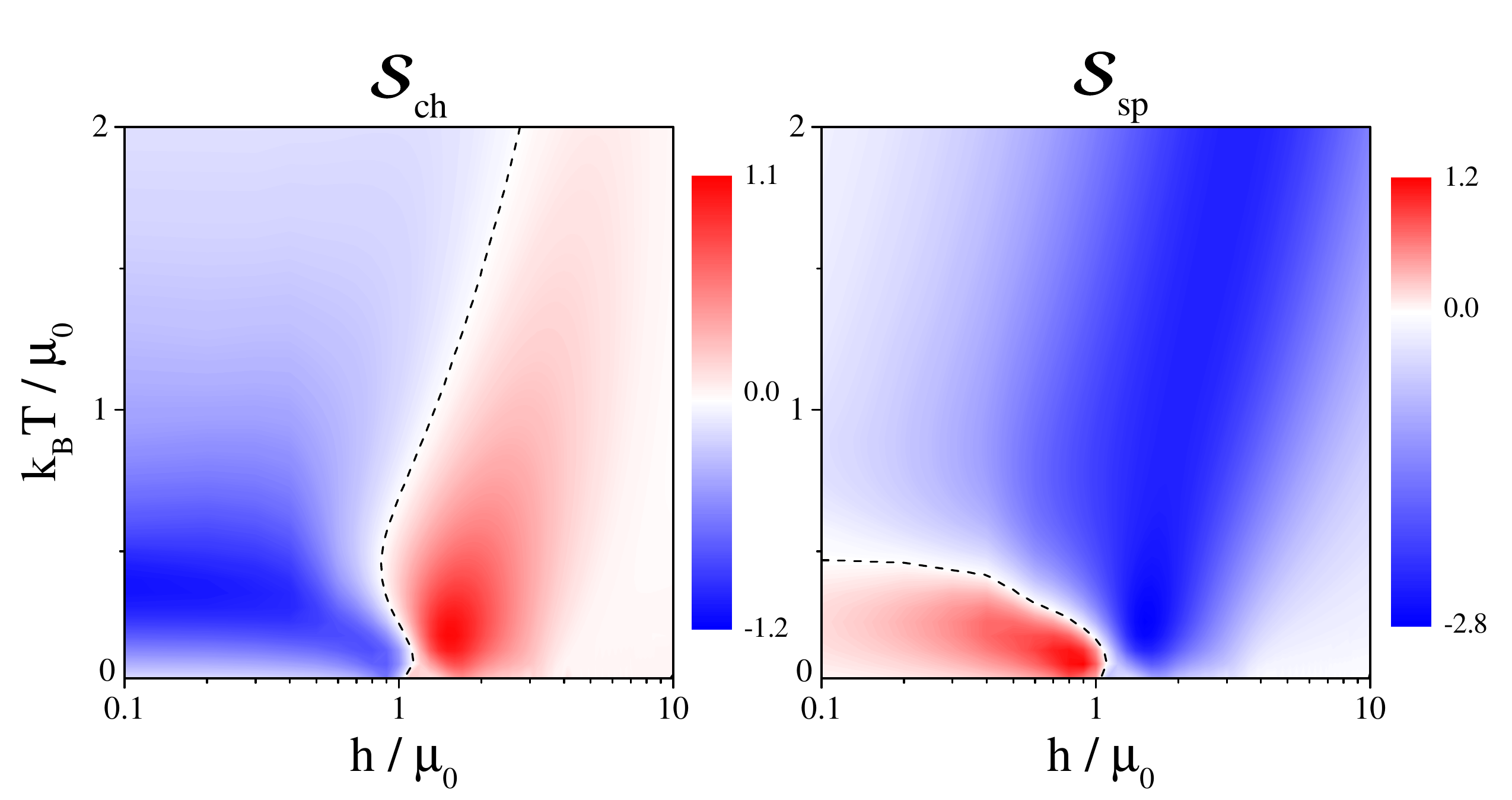}
\caption{\label{fig5}
(Color online) (a) Charge and (b) spin Seebeck coefficients of clean magnetic graphene as functions of normalized exchange field and temperature $k_{B}T/\mu_{0}$ in the ballistic regime. The doping of magnetic graphene $\mu_m$ is assumed to be $1/2$ of the two nonmagnetic leads' doping $\mu_0$. The dashed lines indicate the curves in which the Seebeck coefficients vanish.}
\end{figure}

\subsection{Ballistic transport}
Now we turn to the spin-dependent thermoelectric properties of ballistic graphene. Unlike the diffusive case due to the complicated energy dependence of transmission coefficients we cannot obtain simple analytic relations. Figure \ref{fig5} shows the numerically obtained results for spin and charge thermopowers in ballistic regimes.
Fascinatingly the overall behavior is almost the same as the diffusive regime in the presence of long-range Coulomb impurities. In fact, comparing diffusive and ballistic results, we only see that they are only slightly different in quantitative manner. For instance the possible maxima of thermopowers ${\cal S}_{\rm ch, sp}$ and also the lines in which they vanish are different for two cases which is related to details originating from scattering mechanisms of two regimes. In other words, our results show that despite the details of scattering phenomena, the band structure and dispersion of graphene play a main role in the spin-dependent thermoelectric effects. Of course we know that in the case of diffusive transport the presence of long-range impurities is crucial for thermoelectric effects. Nevertheless when the Seebeck effect does exist, the dependence on the temperature and spin splitting is more or less universal and despite the transport regime we see the same features. It should be noted that such universal behavior is partly related to the definition of Seebeck coefficients themselves and generally we find 
$|{\cal S}|\sim k_B/e$. 
\par
\begin{figure}[tp]
\includegraphics[width=0.95\linewidth]{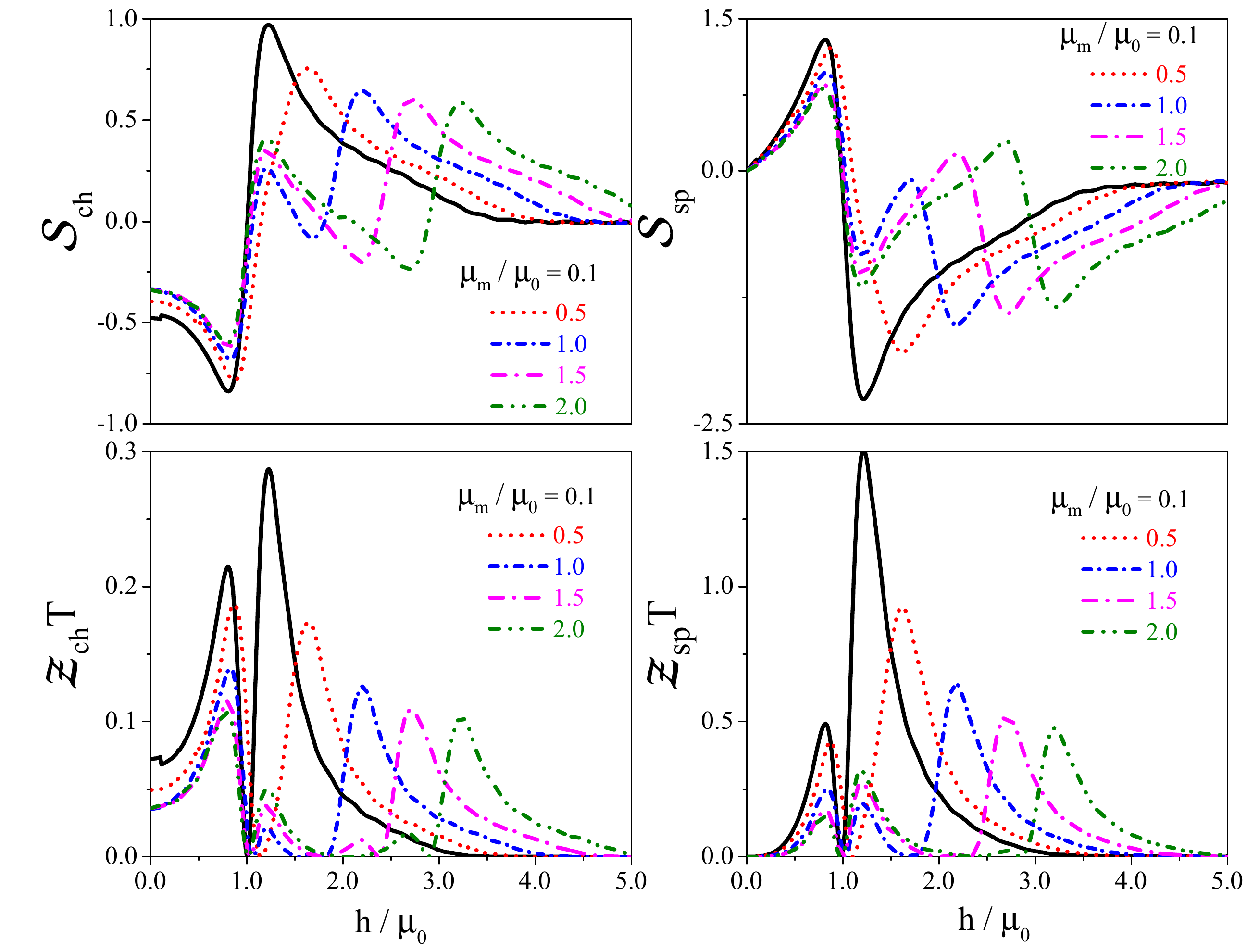}
\caption{\label{fig6} (Color online) The variations of thermopowers and corresponding figures of merit are depicted as functions of normalized exchange splitting for different values of $\mu_{m}/\mu_{0}$. We set the length of the junction $L\mu_{0}/\hbar v_{F}=10.$ and also fix the dimensionless temperature at $k_{B}T/\mu_{0}=0.1$.}
\end{figure}

In the case of ballistic devices we also investigate the effect of gate voltage in the middle region. This could be of great importance in real applications since the gate voltages can be easily tuned. Subsequently one can control the spin-dependent thermoelectric properties by changing the chemical potential $\mu_m$ of middle graphene between the two electrodes and can tune the spin-caloritronic properties. The dependence of spin and charge thermopowers and corresponding figures of merit on the exchange splitting scaled by the leads' chemical potential at zero temperature $\mu_0$ are shown in Fig. \ref{fig6} for a variety of $\mu_m$. First of all we see that changing $\mu_m$ results in shifts in the dependencies of ${\cal S}_{\rm ch, sp}$ and ${\cal Z}_{\rm ch, sp}T$ which can be easily understood due to the fact that the middle magnetic graphene doping plays the main role in transport properties rather than the leads' doping $\mu_0$. Second it is clear that the overall amplitude of the Seebeck coefficients and figures of merit also vary by changing the gate voltages. This is related to the fact that when the chemical potential of electrodes and middle graphene are different the energy-dependent transmission coefficient $T_s(\varepsilon,\phi)$ changes. However unless $\mu_m\ll \mu_0$ these changes do not affect the magnitude of thermopowers since both ${\cal L}_s^0$ and ${\cal L}_s^1$ scale with overall transparency of the scattering region (middle graphene) and only the shift as a function of $V_{\rm ex}/\mu_0$ is observable.

\subsection{On the experimental reliability}
\label{experi}
We close our discussion with commenting on the possible experimental realization of the results we find. First of all we should recall that in all our models we have ignored spin relaxation and spin-flip scattering which is verified until the device size is smaller than spin relaxation length. So in order to have a strong spin Seebeck effect and usage for spin-caloritronics application, we need devices of length $L\lesssim \ell_{sp}\sim 1 \mu m$ which is easily accessible in current experimental devices \cite{wees2014}. On the other side an important step is to combine already existing experimental spintronic and thermoelectric setups based on graphene. This is apparently an easy task since on one side nonlocal magnetoresistance measurements are proven to be very useful to detect spin injection and spin currents in graphene-based spintronic devices. On the other hand the thermoelectric effects themselves have been already observed with significant precision \cite{zuev2009, wei2009} which suggests that spin-dependent thermoelectric properties, in principle, could be detected with high feasibility.  
\par
\begin{figure}[tp]
\includegraphics[width=1.\linewidth]{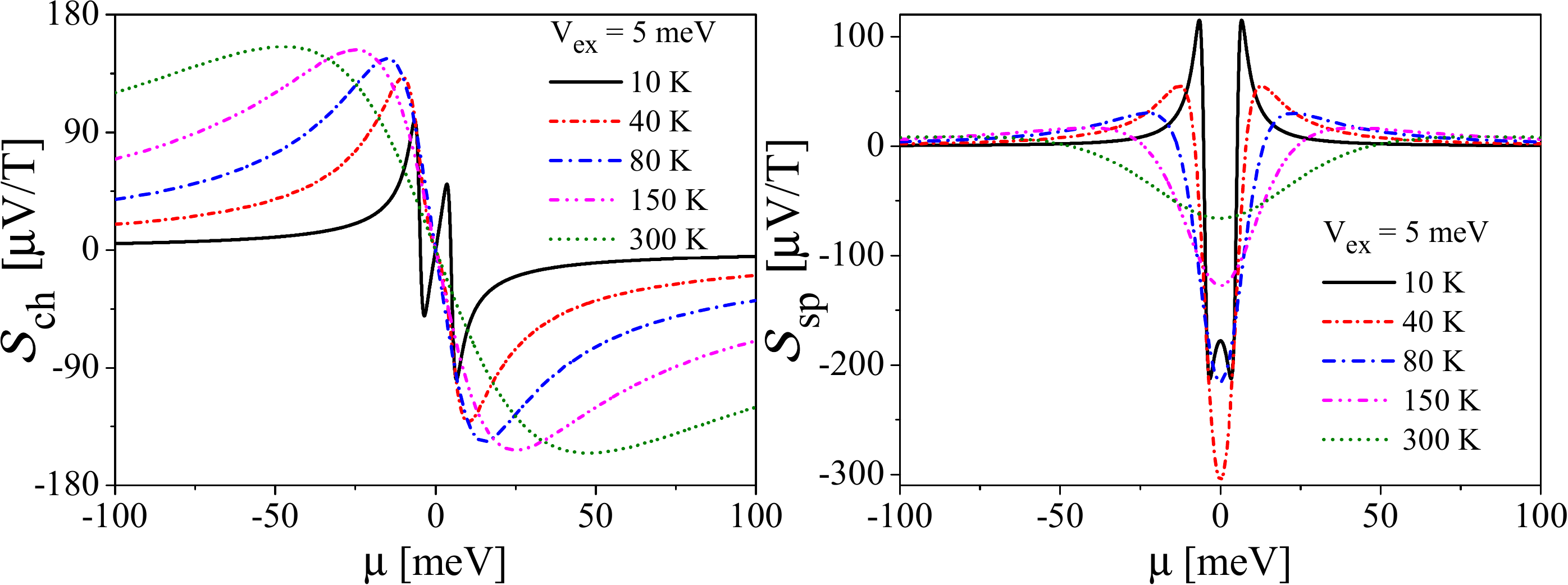}
\caption{\label{fignew} (Color online) The charge and spin thermopower dependence on chemical potential for various temperatures from 10 K to room temperature. The exchange splitting is assumed to be $V_{\rm ex}=5 meV$ in agreement with theoretical prediction of proximity-induced magnetism in graphene.}
\end{figure}

In order to reveal the relevance of our finding to
the experimental situations, in Fig. \ref{fignew} the variations of
charge and spin thermopowers with chemical potential ($\mu$)
for different temperatures are shown in the case of the diffusive regime. In this figure unlike
previous ones instead of scaled dimensionless parameters
we use reliable numerical values of parameters in electronvolts, kelvins, etc., and in particular the temperatures
are exactly the same as in Ref. \onlinecite{zuev2009}. In addition we assume the predicted value for exchange splitting $V_{\rm ex}\sim 5 meV$ \cite{haugen2008}. We see that the numerical values of thermopowers reach values on the order of a hundred $\mu V/T$ which is consistent with previous experimental results for nonmagnetic graphene. In fact when the temperature is large enough in comparison with exchange splitting ($k_BT\gtrsim V_{\rm ex}$) the qualitative behavior of ${\cal S}_{\rm ch}$ is very close to that obtained by Zuev \textit{et al}. \cite{zuev2009}. But very interestingly close to the Dirac point ($\mu=0$) the spin Seebeck coefficient becomes very large especially for intermediate temperatures ($k_BT\sim V_{\rm ex}$) as discussed before.
\par  
Finally, we should comment on the possible influences of phonons in our result which we have not considered. The main effect of phonons is their contribution in the thermal conductivity ${\cal K}$ and the charge and spin thermopower are not affected with the presence of phonons. Therefore it is clear that the thermal conductivity of phonons ${\cal K}_{\rm ph}$ can only affect the figures of merit in our results and since it does not depend on chemical potential or exchange splitting it will only increase ${\cal K}$ depending on temperature. This will decrease ${\cal Z}T$ but dependence on $\mu$ and $V_{\rm ex}$ will not be changed qualitatively, whatsoever. 
Second, the thermal conductivity of phonons will decrease according to some power-law behavior ${\cal K}_{\rm ph} \propto T^{\alpha}$ with $\alpha=1.68$ upon decreasing temperature \cite{xu2014}, while the electrons contribution (${\cal K}_{\rm el}$) in the thermal conductivity varies linearly with $T$ at low temperatures. This can be easily seen from the exact formula which can be obtained in the diffusive regime,
\begin{eqnarray}
{\cal K}_{\rm el}&=&\frac{k_{B}^2T}{2\pi\hbar n_{\rm imp}e^{4}}\left[\frac{14}{15}(\pi k_B T)^{2}\nonumber\right.\\
&+&\left.\sum_{s}\frac{(\mu+sV_{\rm ex})^{2}-(\pi k_B T)^{2}}{3(\mu+sV_{\rm ex})^{2}+(\pi k_B T)^{2}}\left(\mu+sV_{\rm ex}\right)^{2}\right]
\label{k_el}
\end{eqnarray}
with impurity concentration $n_{\rm imp}$.
From the experimental results on phonon thermal conductivity in suspended graphene \cite{seol2010}, we can estimate values on the order of ${\cal K}_{\rm ph}\sim 10^{-1},10^{-2}, 10^{-3} ~\mu W/K$ at temperatures $T\sim 300, 60, 20 \,K$, respectively. Then from Eq. (\ref{k_el}) and assuming typical values $n_{\rm imp}=0.2\times 10^{10} cm^{-2}$, $\mu=0.1 eV$, $V_{\rm ex}=5 meV$ we see that for $T\lesssim 100$ thermal conductivity of electrons decreases linearly with $T$ and for instance at $T=20 K$, we get ${\cal K}_{\rm el}\sim 10^{-9} W/K$ which is the same as the phonon contribution. So we can conclude that for low temperatures $T\lesssim 10 K$ the electrons dominate the thermal conductivity in graphene. At higher temperatures phonons becomes important but as we mentioned before it only results in the overall decline of predicted figures of merit, without affecting their qualitative behavior.

\section{Conclusions}
In this study we reveal that magnetic graphene could be very promising for spin-caloritronics studies and applications. Employing Boltzmann and Landauer formalisms, the spin-dependent thermoelectric properties of graphene in both diffusive and ballistic regimes are obtained. The main finding is that while in the absence of spin splitting, the temperature gradient drives a charge current in graphene, by imposing spin splitting a significant spin current is established, too. 
Very intriguingly when we consider an undoped magnetic graphene in which different spin carriers belong to conduction and valence bands, we will have a pure spin thermopower without charge thermopower. This pure spin current generation by temperature gradient can be achieved in the temperature and spin splitting of the order of the unpolarized state Fermi energy which is accessible in current experiments. So based on this study, we believe that besides the suggested applications of graphene for spintronic devices due to long spin relaxation, magnetic graphene can be used as a base material to investigate spin-thermoelectric phenomena.

\section*{Acknowledgments}
Authors would like to thank Gerrit Bauer for fruitful discussion. A.G.M. acknowledges warm the hospitality of Michele Govenale and Ulrich Z\"ulicke and their useful comments on the manuscript, during his visit at {\color{magenta}Victoria University of Wellington} where part of this work was done. B.Z.R. thanks the CMSP of {\color{magenta} ICTP} in Trieste for hospitality and support during his visit to this institute where part of this work was done. We are also grateful to Saeed Abedinpour for his useful comments.

\appendix

\section{Temperature dependence of the chemical potential}\label{app-a}
\begin{figure}[htp]
\includegraphics[width=0.8\linewidth]{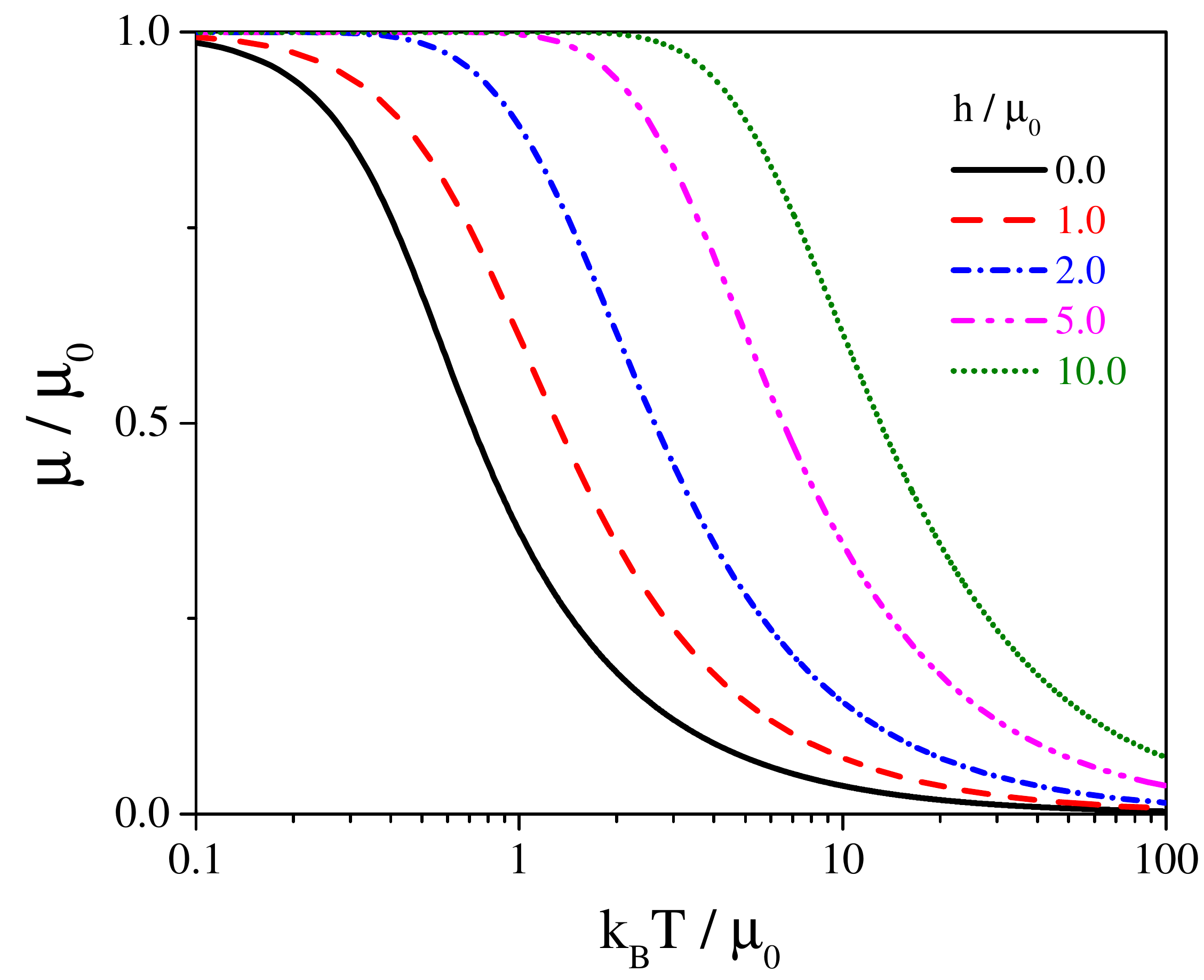}
\caption{\label{fig-app}
(Color online) Temperature dependence of chemical potential $\mu(T)$ for different spin splittings $V_{\rm ex}$. All energies are scaled with respect to $\mu_0$, the chemical potential at zero temperature $T=0$. 
 }
\end{figure}
In this appendix we will present the results of chemical potential $\mu$ variations with temperature $T$ and spin splitting $V_{\rm ex}$. 
The conventional way to obtain the temperature dependence of chemical potential is to enforce the following quantity to be constant,
\begin{eqnarray}
\int d\varepsilon \rho(\varepsilon) f_{\mu(T)}(\varepsilon)={\rm const.}~,
\end{eqnarray}
with Fermi distribution function 
\begin{eqnarray}
f_{\mu(T)}(\varepsilon)=[1+e^{\frac{\varepsilon-\mu(T)}{k_BT} }  ]^{-1}
\end{eqnarray}
which is nothing but the total number of electrons in the system. However in the case of the massless Dirac model for graphene in which there is no lower band for energy the integration over energy diverges. However we can easily overcome this difficulty by subtracting the infinite number of negative energy states. So we define the excess number of charge carriers instead of all electrons, 
\begin{eqnarray}
N_{\rm exc}=\int_{-\infty}^{\infty} d\varepsilon \rho(\varepsilon) f_{\mu(T)}(\varepsilon)-\int_{-\infty}^{0} d\varepsilon \rho(\varepsilon)~,
\end{eqnarray}
which must be a constant irrespective of temperature variations.
Inserting the density of states $\rho(\varepsilon)$ of magnetic graphene and equating the finite-temperature value of the above expression with its zero-temperature correspondence, we will have
\begin{eqnarray}\label{mu-T}
\int_{0}^{\infty} d\varepsilon  \, \varepsilon&&\sum_{s} \left[ f_{\mu(T)}(\varepsilon+sV_{\rm ex})-f_{-\mu(T)}(\varepsilon+sV_{\rm ex})  \right] \nonumber \\
= &&\frac{1}{2}\sum_{s} (\mu_0+sV_{\rm ex})^2 {\rm sgn}(\mu_0+sV_{\rm ex})    ~,
\end{eqnarray}

It is worth noting that here the chemical potential will depend on $V_{\rm ex}$ as well. Scaling all the energies with zero-temperature doping $\mu_0$, by solving Eq. (\ref{mu-T}) numerically we could obtain $\mu(T)/\mu_0$ as a function of normalized temperature $k_BT/\mu_0$ and spin splitting $V_{\rm ex}/\mu_0$. The results can be seen in Fig. \ref{fig-app}.

\bibliography{grph-sbk}

\end{document}